\documentclass[conference,letterpaper]{IEEEtran}
\usepackage[utf8]{inputenc} 
\usepackage[T1]{fontenc}
\usepackage{url}
\usepackage{ifthen}
\usepackage{cite}
\usepackage{amsmath,amssymb,amsfonts}
\usepackage{amsthm}
\usepackage[inline]{enumitem}
\usepackage{textcomp}
\usepackage{xcolor}
\usepackage{float}
\def\BibTeX{{\rm B\kern-.05em{\sc i\kern-.025em b}\kern-.08em
    T\kern-.1667em\lower.7ex\hbox{E}\kern-.125emX}}

\newtheorem{auxiliary code}{Auxiliary Code}

\usepackage{algorithm}
\usepackage[noend]{algpseudocode}
\algrenewcommand\algorithmicindent{0.8em}%
\usepackage{tikz}
\usepackage{graphicx}
\usetikzlibrary{positioning}
\usepackage[font={small}]{caption}
\usepackage{courier}
\usepackage{subcaption}
\usepackage{multirow}
\usepackage[T1]{fontenc}    
  \usepackage{array}  

\newcolumntype{L}{>{\centering\arraybackslash}m{2.5cm}}    
\newcolumntype{M}{>{\centering\arraybackslash}m{4cm}} 
\newcolumntype{N}{>{\centering\arraybackslash}m{5cm}} 
\newcolumntype{O}{>{\centering\arraybackslash}m{5cm}} 
    
\DeclareCaptionLabelFormat{andtable}{\tablename~\thetable  \hspace{0.1cm}\& #1~#2   }    
    
\def \tGF {\mathbb{GF}}
\def \tX {\mathbf{X}}
\def \tY {\mathbf{Y}}

\def \tH {\mathbf{H}}
\def \tRm {\mathbf{S}}
\def \tbs {\alpha}
\def \tch {\gamma}
\def \tL {L}
\def \tR {P}
\def \tDiffMutation {{\fontfamily{qcr}\selectfont DiffMutation}}
\def \tCrossOver {{\fontfamily{qcr}\selectfont CrossOver}}

\newcommand\deb[1]{{\color{black}#1}}    

\usepackage{accents}

\begin{document}

\title{Non-Binary LDPC Code Design for Energy-Time Entanglement Quantum Key Distribution\vspace{-0.75cm}}

\author{\IEEEauthorblockN{Debarnab Mitra, Lev Tauz,  Murat Can Sarihan, Chee Wei Wong, and Lara Dolecek}
\IEEEauthorblockA{Department of Electrical and Computer Engineering, University of California, Los Angeles, USA\\
email: debarnabucla@ucla.edu, levtauz@ucla.edu, mcansarihan@ucla.edu, cheewei.wong@ucla.edu, and   dolecek@ee.ucla.edu}\vspace{-1.2cm}}

\maketitle

\begin{abstract}

In energy-time entanglement Quantum Key Distribution (QKD), two users extract a shared secret key from the arrival times (discretized as symbols) of entangled photon pairs. In prior work,  Zhou \emph{et al.} proposed a \emph{multi-level coding} (MLC) scheme that splits the observed symbols into bit layers and utilizes binary Low-Density Parity-Check (LDPC) codes for \emph{reconciliation} of the symbols. While binary LDPC codes offer low latency for key generation, splitting the symbols into bits results in a loss of key generation rate due to error propagation. Additionally, existing LDPC codes do not fully utilize the properties of the QKD channel to optimize the key rates. In this paper, we mitigate the above issues by first generalizing the MLC scheme to a non-binary(NB) MLC scheme that 
has layers with non-binary symbols and utilizes NB-LDPC codes. We show the NB-MLC scheme offers flexibility in system design. Additionally, we show that the NB-MLC scheme with a small symbol size per layer offers the best trade-off between latency and key rate. We then propose a framework to jointly optimize the rate and degree profile of the NB-LDPC codes that is tailored towards the QKD channel resulting in higher key rates than prior work.
\end{abstract}

\vspace{-0.16cm}
\section{Introduction}\label{sec:intro}
\vspace{-0.09cm}

Quantum Key Distribution (QKD) provides a physically secure way to share a secret key between two users, Alice and Bob, over a quantum communication channel in the presence of an eavesdropper Eve \cite{Original-QKD, QKD-nature, ET-QKD-1}. Energy-time entanglement QKD (ET-QKD) protocols have been studied extensively in literature due to their ability to extract multiple bits per generated entangled photon pairs \cite{ET-QKD-1,Murat}. At a high level, an ET-QKD protocol consists of the following steps \cite{MLC}: i) In the first step, called \emph{generation}, Alice and Bob generate \emph{raw keys} using a quantum channel that are represented as sequences of symbols. Due to imperfections in the quantum channel, the raw keys at Alice and Bob may disagree in some positions; ii) In the second step, called \emph{information reconciliation} (IR), Alice and Bob communicate over a public channel (accessible to Eve) to reconcile the raw keys; iii) In the third step, called \emph{privacy amplification} (PA), Alice and Bob amplify the privacy of the reconciled key by accounting for Eve's knowledge to generate the final shared secret key. Channel coding is utilized in the IR step to ensure that Alice and Bob arrive at an identical sequence of symbols. In this paper, similar to \cite{MLC}, we focus on the IR step of the protocol and assume that if we communicate $m$ bits during the IR step, then PA results in a loss of $m$ bits from the length of the reconciled key to get the shared secret key. The \emph{key rate} of the system is defined as the average length of the shared secret key obtained by Alice and Bob after PA.

A promising coding framework proposed to get high key rates is called the \emph{multi-level coding} (MLC) scheme \cite{MLC} that has been considered for works such as \cite{ET-QKD-1, MLC-application}. In the MLC scheme, the sequence of symbols after the generation step is converted into multiple bit layers and then each bit layer is sequentially reconciled using binary LDPC codes.
Binary LDPC codes have low complexity and fast decoding algorithms and hence result in low latency and complexity for key generation. However, the MLC coding scheme suffers from error propagation where a decoding error in one of the bit layers results in decoding errors in subsequent bit layers leading to \deb{reduced} key rates. Contrary to the MLC scheme with binary LDPC codes, non-binary (NB) LDPC codes that directly encode the generated symbols do not suffer from error propagation. Hence, NB-LDPC code can naturally lead to higher key rates. However, the symbols in the generation step can belong to a Galois field of size as large as $2^{10}$ and it is known that iterative decoding of NB-LDPC codes has a very high complexity (log-linear in the field size\cite{Mackay}) leading to high latency for the key generation. Hence, \deb{baseline} NB-LDPC codes with large field sizes are not favorable in QKD applications requiring low latency, such as in \cite{latency-1, latency-2}. 

In addition to the above latency vs. key rate trade-off, the LDPC codes used previously in the IR step of ET-QKD protocols have not fully utilized the properties of the ET-QKD channel. For example, 
\cite{MLC} used a standard LDPC ensemble without optimization. Similarly, spatially-coupled (SC) LDPC codes, irregular repeat accumulate (IRA) codes, SC-IRA codes, and multi-edge-type (MET) codes have been discussed for the continuous-variable (CV) QKD \cite{SC-QKD, IRA-QKD}. However, these works focus on channel models such as binary input additive white Gaussian noise (BIAWGN) that do not match the ET-QKD channel \cite{PriscaAsilomar}.

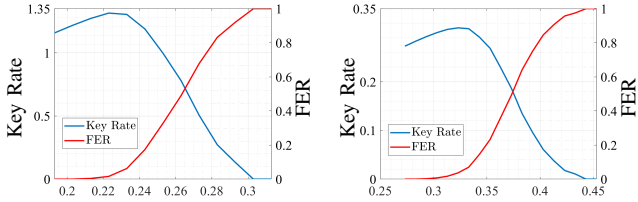
\begin{figure}[t]
    \centering
    \begin{subfigure}{0.5\linewidth}
\begin{minipage}{0.99\linewidth}
\begin{tikzpicture}
  \node (img) {\includegraphics[scale=0.125]{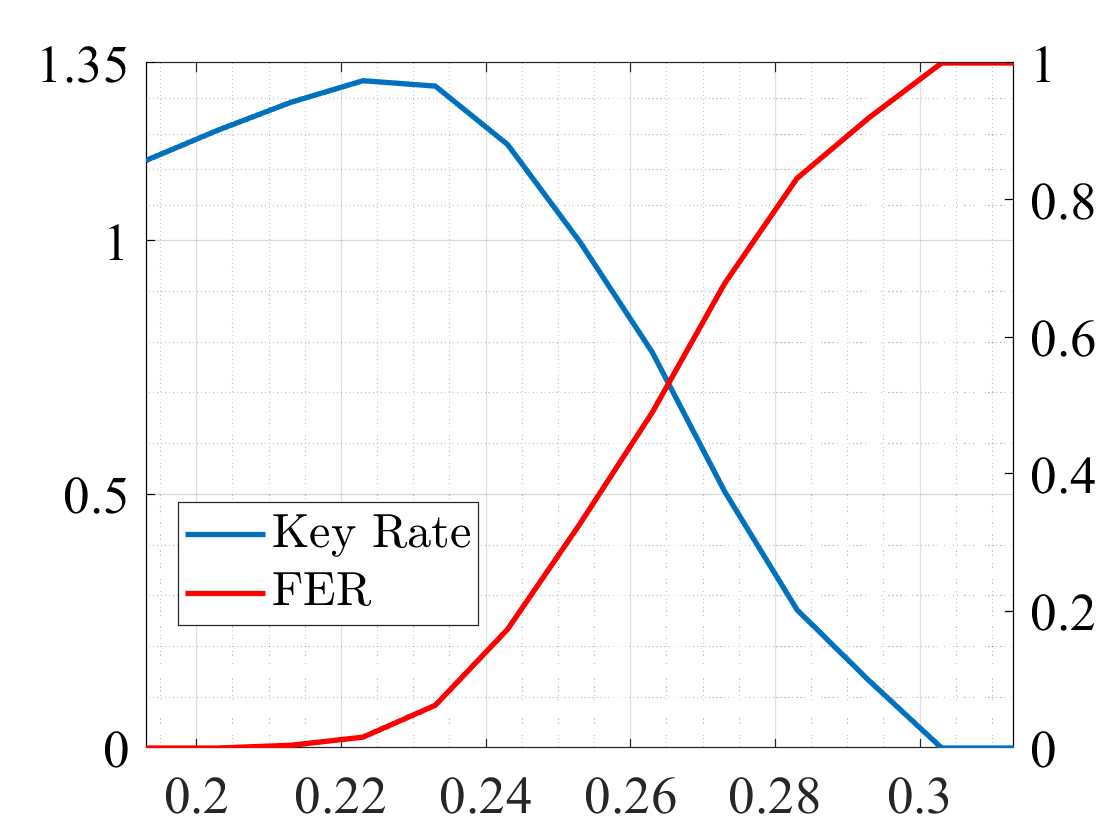}};
  \node[below=of img, node distance=0cm, yshift=1.1cm,font=\color{black}] {\footnotesize{Coding Rate}};
  \node[left=of img, node distance=0cm, rotate=90, anchor=center,yshift=-1.1cm,font=\color{black}] {\footnotesize{Key Rate}};
  \node[right=of img, node distance=0cm, rotate=90, anchor=center,yshift=1.05cm,font=\color{black}] {\footnotesize{FER}};
 \end{tikzpicture}
 \end{minipage}
 \vspace{-3pt}
    \end{subfigure}%
    \begin{subfigure}{0.5\linewidth}
\begin{minipage}{0.99\linewidth}
\begin{tikzpicture}
  \node (img) {\includegraphics[scale=0.125]{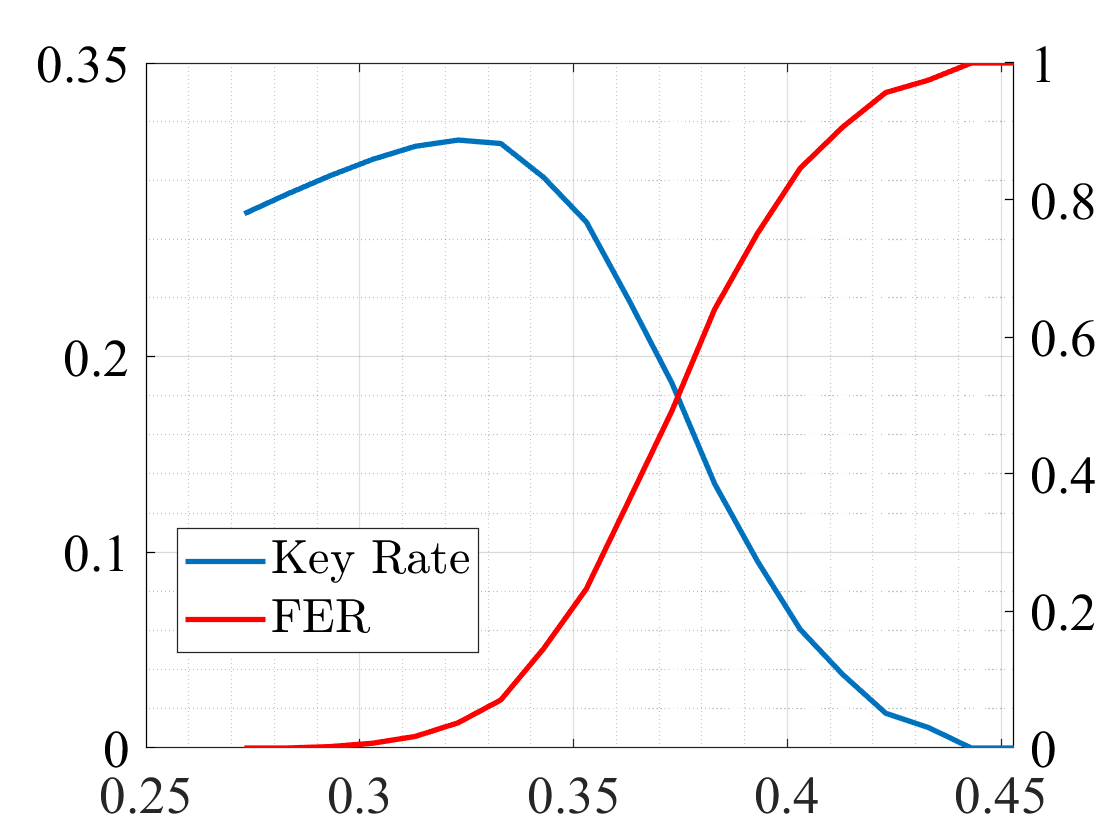}};
  \node[below=of img, node distance=0cm, yshift=1.1cm,font=\color{black}] {\footnotesize{Coding Rate}};
  \node[left=of img, node distance=0cm, rotate=90, anchor=center,yshift=-1.2cm,font=\color{black}] {\footnotesize{Key Rate}};
  \node[right=of img, node distance=0cm, rotate=90, anchor=center,yshift=1.05cm,font=\color{black}] {\footnotesize{FER}};
 \end{tikzpicture}
 \end{minipage}
 \vspace{-3pt}
    \end{subfigure}
     \vspace{-15pt}
    \caption{Key rate and FER vs.~coding rate.
    Left panel: NB-LDPC code in $\tGF(2^6)$; Right panel: Binary LDPC code. Maximum in the key rate occurs at FER around 0.05 in both figures. }
    \label{fig:motivation}
\end{figure}

A unique property of the ET-QKD problem considered in this paper is that the key rate of the system is closely dependent on both the rate of the code and the frame error rate (FER) performance. Fig. \ref{fig:motivation} shows the FER and key rates obtained by a random LDPC code for different values of rate. From this graph, we see that increasing the code rate can improve the key rate even at the cost of higher FER, a phenomenon we see in both binary and non-binary LDPC codes. Additionally, the maximum in the key rate occurs for a relatively large value of FER ($\sim 5$\%). 
While the conventional code design approach is to minimize the FER to a very small value for a given rate, in this case, the goal is to jointly optimize both the rate and the FER to achieve the largest key rate. 

The degree distribution of an LDPC code is known to affect its FER performance. Degree distribution optimization techniques for LDPC codes based on code thresholds (e.g., \cite{DegreeDist}) optimize the degree distribution for a fixed rate and hence are not directly applicable to the current ET-QKD problem that needs a joint rate and FER optimization. Additionally, the optimized degree distributions are designed for non-QKD channels (e.g., BIAWGN in \cite{DegreeDist}) and they do not result in large key rates as we demonstrate in Section \ref{sec:sims}.

\begin{figure}[t]
\centering
\includegraphics[scale=0.3]{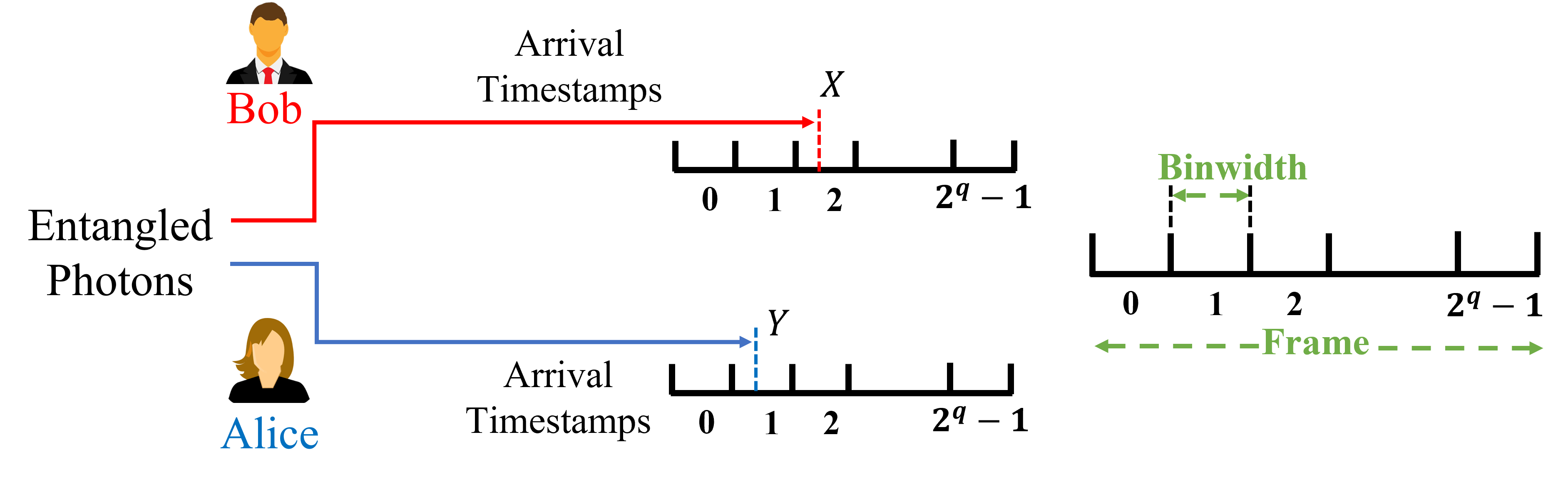}
     \vspace{-5pt}
    \caption{QKD system model. The arrival times of photons are discretized using pulse position modulation. Each frame has $2^q$ bins and the spacing between frames in called binwidth.}
    \label{fig:system_model}
    \vspace{-15pt}
\end{figure}

In the paper, we mitigate the above issues of latency vs. key rate trade-off and code design considering the properties of the ET-QKD channel using a two-pronged approach. 
Firstly, we  generalize the MLC scheme of \cite{MLC} to
a non-binary MLC scheme by splitting the symbols after the generation step into multiple layers with non-binary symbols belonging to a smaller Galois field. The NB-MLC scheme offers a natural trade-off between latency and key rate depending on the size of the symbols in a layer, allowing flexibility in system design. Additionally, we demonstrate that the NB-MLC scheme with a small symbol size per layer results in higher key rates compared to a fully binary scheme \cite{MLC} as well as using a fully non-binary scheme without layering. 
Secondly, we provide a joint rate and degree distribution optimization (JRDO) framework based on differential evolution \cite{DiffEvol} for the construction of the NB-LDPC codes in each layer of the NB-MLC scheme. The JRDO framework uses the QKD channel information and we demonstrate that it results in a higher key rate compared to the LDPC codes used in the MLC scheme \cite{MLC} and that obtained by utilizing degree distributions optimized for conventional channels such as the BIAWGN channel \cite{DegreeDist}.
The rest of this paper is organized as follows.  In section \ref{sec:prelims}, we provide the preliminaries and the system model. In section \ref{sec:NB-MLC}, we describe the NB-MLC scheme. In section \ref{sec:JointOpt}, we provide the JRDO framework. Finally, we provide simulation results in section \ref{sec:sims} and conclude the paper in section \ref{sec:conclusion}. 

\vspace{-0.2cm}
\section{Preliminaries and System Model}\label{sec:prelims}
\vspace{-0.1cm}

\subsubsection{ET-QKD system model} As shown in Fig. \ref{fig:system_model}, in ET-QKD \cite{ET-QKD-1}, energy-time entangled photon pairs are generated by a third party in the generation step. Alice and Bob then receive one photon each out of the pair who then record the arrival times of the received photons. The raw key information is derived from the arrival times. In this method, Alice and Bob both synchronize their timelines and then discretize their time into frames where each frame is further divided into $2^{q}$ bins of equal size, \deb{where $q$ is a positive integer}. Alice and Bob retain
only time frames in which they both detect a single photon arrival and discard all other frames. The photon arrival time in a non-discarded frame is then converted (discretized) into a symbol in $\tGF(2^{q})$ based on the bin number the received photon occupies within each frame. The discretized sequences received by Alice and Bob are then divided into blocks each having $N$ symbols where $N$ is the code length. Let $\tX = \{X_1, \ldots, X_N\}$, $X_i\in \tGF(2^q)$ and $\tY = \{Y_1, \ldots, Y_N\}$, $Y_i \in \tGF(2^q)$ be the sequences of length $N$ recorded by Alice and Bob, respectively. 
Due to imperfections in the generation step (e.g., timing jitters, transmission loss \cite{MLC}) $\tY$ is a noisy version of $\tX$. We assume the sequences $\tX$ and $\tY$ are memoryless and each $Y_i$ is the output of the ET-QKD channel characterized by transition law $P_{Y|X}$ and input $X_i$. 

A simple IR protocol based on NB-LDPC codes in $\tGF(2^q)$ proceeds as follows. Alice sends Bob $\tRm = \tH\tX$ over the public channel (which is accessible to Eve) where $\tH \in \tGF(2^q)^{M\times N}$ is the parity check matrix of an NB-LDPC code. Bob decodes $\tX$ using the received $\tRm$ and side information $\tY$. LDPC decoding using side information is encountered in the Slepian-Wolf (SW) problem \cite{SW-decoding}. SW LDPC decoding is very similar to the sum-product decoder used in conventional decoding of LDPC channel codes with small differences in the way the log-likelihood messages are initialized and the CN to VN messages. We refer the reader to \cite{SW-decoding} for details about SW LDPC decoding. The goal of the NB-LDPC code is to make the decoding output equal to $\tX$ with high probability while  ensuring that the information leaked to Eve is minimized. Finally, the sequence $\tX$ is the reconciled key that is passed to the PA step. The key rate $r$ (in bits per photon) of the above scheme (similar to \cite{MLC}) is given as follows:
\vspace{-0.1cm}
\begin{equation}\label{eqn:key_rate_nb}
    r = q(1 - E)\frac{N-M}{N},
\end{equation}
where $E$ is the FER incorporated in the decoding of $\tX$. Note that we subtract $M$ in Eqn. \eqref{eqn:key_rate_nb} since $M$ symbols are sent over the public channel and hence will be lost due to PA.

\subsubsection{ET-QKD channel} \label{sec:qkd_channel}
In this paper, we use empirical data from a practical \deb{ET-QKD system testbed\cite{Murat}} to estimate the channel transition law $P_{Y|X}$ directly. For interested readers, \deb{authors in \cite{PriscaAsilomar}} have demonstrated a modeling that provides a good approximation of the ET-QKD channel. Succinctly, the ET-QKD channel
is a mixture of \emph{local} and a \emph{global} channel with Gaussian and uniform distributions. The uniform distribution causes a low 
SNR in our system resulting in a high operating FER ($\sim1-10 \%$). Note that the ET-QKD channel is different from conventional channels such as AWGN, BSC, etc. As such, codes that have been optimized for these channels are not \deb{necessarily} the best ones for the ET-QKD channel as we demonstrate in Section \ref{sec:sims}.

\subsubsection{NB-LDPC codes}\label{sec:ldpc_prelims}
A NB-LDPC code over $\tGF(2^q)$ is defined by a sparse parity check matrix $\tH \in \tGF(2^q)^{M \times N}$. The matrix $\tH$ has a Tanner graph representation comprising of $M$ check nodes (CNs) and $N$ variable nodes (VNs) corresponding to rows and columns of $\tH$. A CN is connected to a VN by an edge if the corresponding entry in $\tH$ is non-zero where the edge is additionally labeled by the non-zero entry. The interconnection between VNs and CNs of a code is represented by degree distributions $\tL(x) = \sum_d \tL_dx^{d}$ and $\tR(x) = \sum_d\tR_dx^{d}$, where $\tL_d$ and $\tR_d$ represent the fraction of nodes connected respectively to VNs and CNs of degree $d$. The coding rate $R$ of the code is given by 
$R = 1 - \frac{\tL'(1)}{\tR'(1)}$. 
The FER performance of the code depends on the degree distributions  $\tL(x)$ and $\tR(x)$. 
In this paper, we optimize the rate $R$ and VN degree distribution $\tL(x)$. For given $R$ and $\tL(x)$, we find a two-element distribution $\tR(x)$ that results in rate $R$. Now for the  degree distributions $\tL(x)$ and $\tR(x)$, parity check matrix $\tH$ is randomly sampled among the ensemble of LDPC codes that match these degree distributions \cite{ModernCodingTheory} and label each edge uniformly at random with a non-zero element of $\tGF(2^q)$. In the next section, we propose the NB-MLC scheme for IR.

\vspace{-0.15cm}
\section{Non-Binary Multi-Level Coding}\label{sec:NB-MLC}
\vspace{-0.1cm}

The NB-MLC scheme offers a tradeoff between key rate $r$ and latency/complexity of key generation through an integer parameter $a$, $1 \leq a \leq q$. Let $b$ and $r$ be integers such that $q = ab + r$, where $b = \lfloor \frac{q}{a} \rfloor$ and $r$ is the remainder when $q$ is divided by $a$. Each symbol $X$ in $\tX$ received by Alice is an element of $\tGF(2^q)$. We split $X$ into $b+1$ symbols $(X_1, X_2, \ldots X_{b+1})$, where $X_i \in \tGF(2^a), 1 \leq i \leq b$ and $X_{b+1} \in \tGF(2^r)$ using an injective mapping $u: \tGF(2^q) \rightarrow \tGF(2^a)^{b} \times \tGF(2^r)$. Using the above conversion, we split the sequence $\tX$ into $b+1$ layers $(\tX_1, \tX_2, \ldots, \tX_{b+1})$, where $\tX_i \in \tGF(2^a)^{N}, 1 \leq i \leq b$ and $\tX_{b+1} \in \tGF(2^r)^N$. Let $\tbs_i$ denote the bit size of the symbols in the $i$th layer. We have $\tbs_i = a, 1 \leq i \leq b$ and $\tbs_{b+1} = r$.
For each layer $i$, we use a NB-LDPC code $\tH_i$ where $\tH_i \in \tGF(2^{\tbs_i})^{m_i \times N}$ for $1\leq i \leq b+1$. Now, Alice generates a message $\tRm  = \{\tRm_1, \tRm_2, \ldots, \tRm_{b+1}\}$ by setting $\tRm_i = \tH_i\tX_i$, $1\leq i \leq b+1$ and sends it to Bob over the public channel. Using $\tRm$ and $\tY$, Bob decodes every layer $\tX_i, 1\leq i\leq b+1$ and hence $\tX$ which is the reconciled key. 

Let $\widehat{\tX}_1^{i-1}:= \{\widehat{\tX}_1, \widehat{\tX}_2, \ldots, \widehat{\tX}_{i-1}\}$ be the decoding result of layers $1, 2, \ldots i-1$. Similar to \cite{MLC}, Bob decodes layer $\tX_i$ with received message $\tRm_i$, and side information $\tY$ and $\widehat{\tX}_1^{i-1}$ using SW decoding for NB-LDPC codes \cite{SW-decoding}. The equivalent channel for the $i$th layer takes input $\tX_i$ and outputs $\{\tY,\tX_1^{i-1}\}$ with transition law $\tch^{i} := P(Y = y, X^{i-1}_1 = x^{i-1}_1 | X_i = x_i)$. We derive the transition law $\tch^{i}$ empirically from our QKD  testbed and use it in Section \ref{sec:JointOpt} for code optimization. 

The size of massage $\tRm_i$ sent by Alice for the reconciliation of the $i$th layer is $m_i$. Let $E_i$ be the FER for the $i$th layer. 
The key rate of the NB-MLC scheme is obtained by adding the key rates of each layer (similar to \cite{MLC}) and is given by 

\vspace{-0.25cm}
\begin{equation}\label{eqn:key_rate_mlc}
    r = \sum_{i = 1}^{b+1} \tbs_i(1 - E_i)\frac{N-m_i}{N}.
\end{equation}
\vspace{-0.35cm}

The key rate depends on the coding rates $R_i = \frac{N-m_i}{N}$ of $\tH_i$ used in layer $i$. The parameter $a$ in the NB-MLC scheme affects the key rate as well as the latency and hardware complexity (which are, respectively, the sum of the decoding latencies and the sum of the complexities of all the layers). Note that $a = 1$ gives us the binary MLC scheme of \cite{MLC} and $a = q$ provides a completely non-binary scheme with only one layer. As $a$ is increased from $1$ to $q$, the complexity monotonically increases. However, as we demonstrate in Section \ref{sec:sims}, the key rates are not monotonic in $a$.

Finally, the performance of the system in terms of the key rate and latency also depends on the mapping $u(X)$ used to split the symbols $X \in \tGF(2^q)$ into symbols of different layers. For convenience, we use the following mapping. We first convert $X$ into its binary representation $X_b$. We then split the bits in $X_b$ into $b+1$ groups with the $i$th group having $\tbs_i$ bits. We then treat the bits in each group $i$ as a binary representation and convert them back to a symbol in $\tGF(2^{\tbs_i})$. A study on the effects of different mappings on the key rate and latency is beyond the scope of this paper and is part of future work. In the next section, we provide the JRDO framework based on differential evolution to jointly optimize the rate and degree distribution of the NB-LDPC codes.

\vspace{-0.30cm}
\section{Joint Rate and\\ Degree Distribution Optimization}\label{sec:JointOpt}
\vspace{-0.1cm}
In this section, we provide the framework to design parity check matrices $\tH_i$, $1 \leq i \leq b+1$ for use in the $i$th layer of the NB-MLC scheme with channel transition probability $\tch^{i} := P(Y = y, X^{i-1}_1 = x^{i-1}_1 | X_i = x_i)$. 
The construction method is the same for all layers, hence we drop index $i$. 
In particular, we design the VN degree distribution $\tL(x)$ and rate $R$ for $\tH$ (see Section \ref{sec:ldpc_prelims} for how a non-binary $\tH$ is generated from $\tL(x)$ and $R$). The channel transition probability is $\tch$. 

\begin{figure*}[t]
    \centering
    \begin{subfigure}{0.33\linewidth}
\begin{minipage}{0.99\linewidth}
\begin{tikzpicture}
  \node (img) {\includegraphics[scale=0.2]{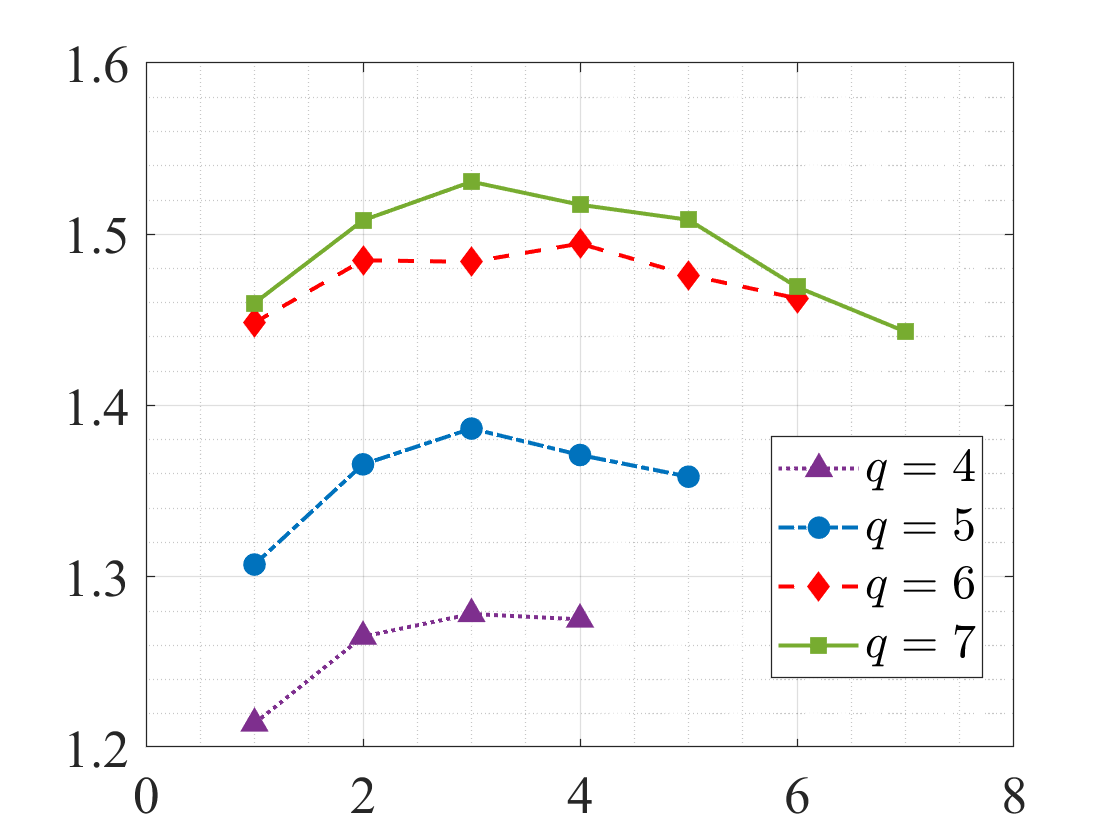}};
  \node[below=of img, node distance=0cm, yshift=1.2cm,font=\color{black}] {$a$};
  \node[left=of img, node distance=0cm, rotate=90, anchor=center,yshift=-1.2cm,font=\color{black}] {Key Rate};
 \end{tikzpicture}
 \end{minipage}
    \end{subfigure}%
    \begin{subfigure}{0.33\linewidth}
\begin{minipage}{0.99\linewidth}
\begin{tikzpicture}
  \node (img) {\includegraphics[scale=0.2]{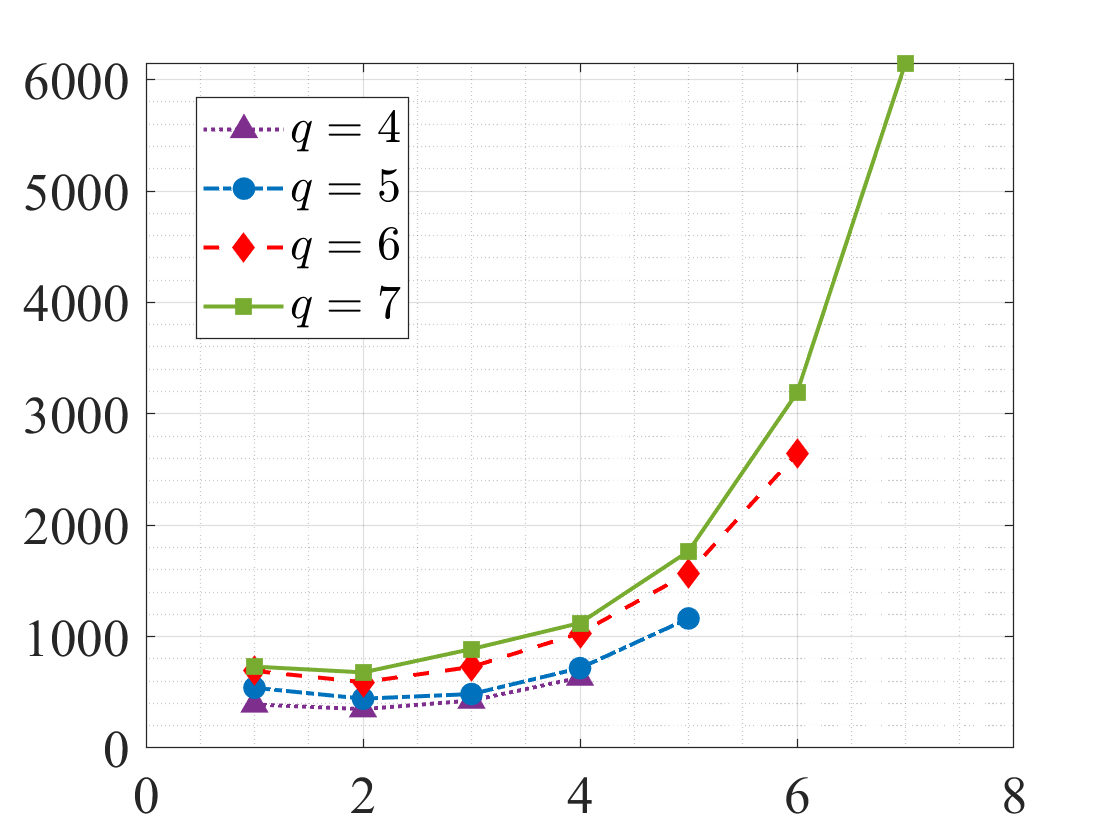}};
  \node[below=of img, node distance=0cm, yshift=1.2cm,font=\color{black}] {$a$};
  \node[left=of img, node distance=0cm, rotate=90, anchor=center,yshift=-1.0cm,font=\color{black}] {Latency (in ms)};
 \end{tikzpicture}
 \end{minipage}
    \end{subfigure}%
        \begin{subfigure}{0.33\linewidth}
\begin{minipage}{0.99\linewidth}
\begin{tikzpicture}
  \node (img) {\includegraphics[scale=0.2]{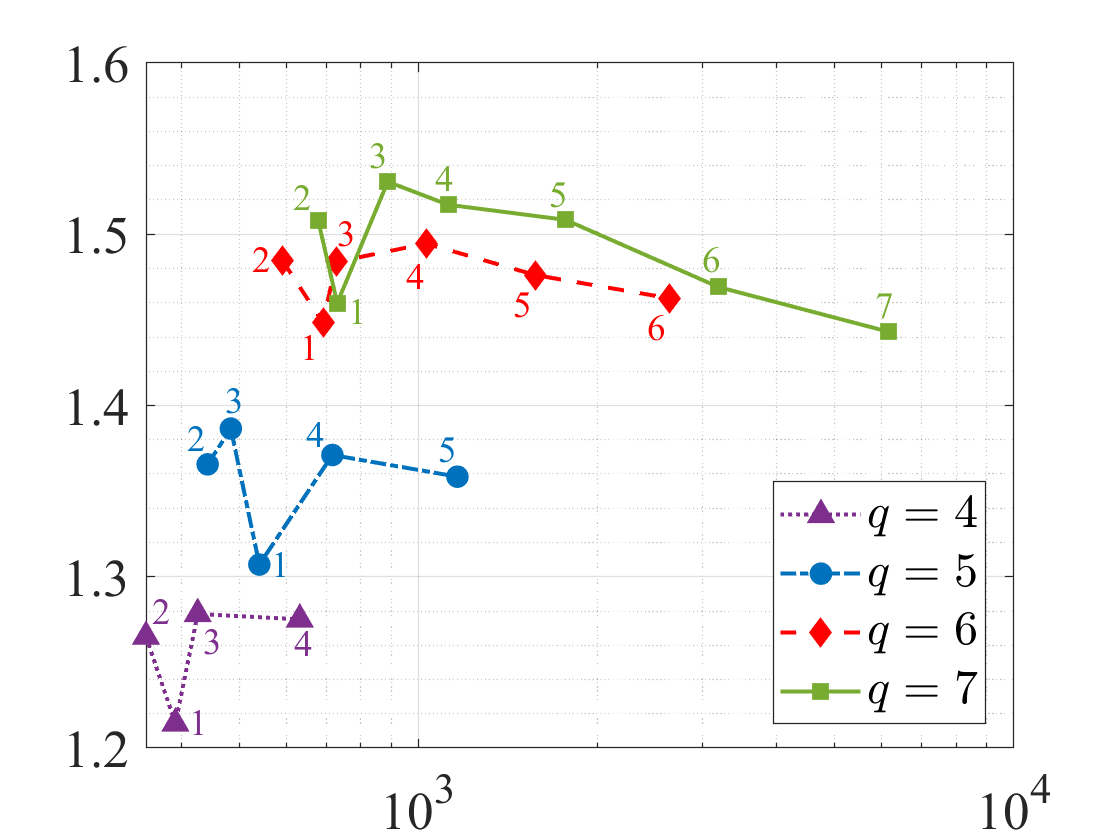}};
  \node[below=of img, node distance=0cm, yshift=1.2cm,font=\color{black}] {Latency (in ms)};
  \node[left=of img, node distance=0cm, rotate=90, anchor=center,yshift=-1.2cm,font=\color{black}] {Key Rate};
 \end{tikzpicture}
 \end{minipage}
    \end{subfigure}
     \vspace{-10pt}
    \caption{Key rate and latency for different $q$ as the NB-MLC bit size $a$ is varied. The ET-QKD system has a binwidth of 300ps. Left panel: Key rate vs. $a$; Middle panel: Latency vs. $a$; \deb{Right panel: Key rate vs. latency where each point on a curve for a particular $q$ represents a different value of $a$ (the values of $a$ are marked on the curves)}. All curves use LDPC codes mentioned in Section \ref{sec:ldpc_prelims} with $L(x) = x^3$. }
    \label{fig:a_Varied}
    \vspace{-15pt}
\end{figure*}
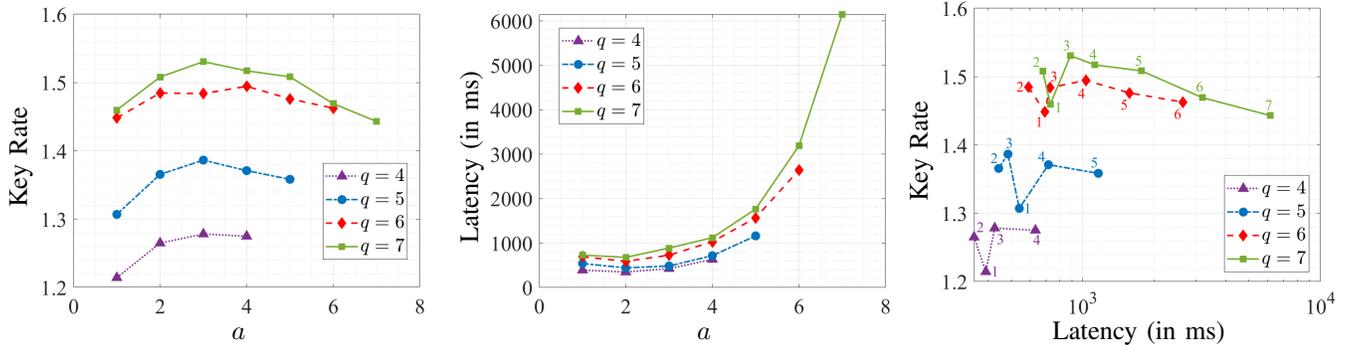

Our framework utilizes differential evolution (DE) \cite{DiffEvol} to find $\tL(x)$ and $R$. DE is a popular and effective population-based evolutionary algorithm that can be used for a maximization (or minimization) of any function $f()$. The algorithm iteratively improves a candidate solution (that maximizes $f()$) using an evolutionary process and can explore large design spaces with low complexity. DE has been extensively used in coding theory literature to design good irregular LDPC codes for the erasure channel \cite{DE-LDPC}, AWGN channel \cite{DegreeDist}, Rayleigh fading channel \cite{DE-LDPC-2}, etc. 
The goal in these works is to design degree distributions that have low FER. This goal is achieved using DE where the function $f()$ is generally set as some low complexity predictor of the FER performance of the code such as the threshold obtained by density evolution \cite{DegreeDist}. However, as discussed in section \ref{sec:intro}, the goal for us in this paper is to maximize the key rate and not \deb{merely to} minimize the FER. Additionally, the techniques for optimizing the degree distributions using code thresholds work for a fixed code rate and we have not found any previous work that jointly optimizes the code rate along with maximizing the threshold.

In this paper, following the expression for key rate in Eqn. \eqref{eqn:key_rate_mlc}, we jointly optimize the degree distribution $\tL(x)$ and the coding rate $R$ using DE by setting $f(\tL(x), R) = (1 - E)R$. Here, $E$ is the expected FER of a code ensemble with degree distribution $\tL(x)$ and rate $R$ on a channel with transition law $\tch$.  Note that to be able to optimize the above function using DE feasibly, the cost of computing the function must be low (since the DE algorithm evaluates the function $f()$ a certain fixed number of times at every iteration). However, as discussed in Section \ref{sec:qkd_channel}, since the FER of the code is high ($\sim 1-10\%$), the FER $E$ can be easily computed using Monte-Carlo (MC) simulations with a small number of MC experiments (e.g., 200-300).  The overall JRDO algorithm is provided in Algorithm \ref{alg:DE_opt} where the procedures \tDiffMutation() and \tCrossOver() have regular meanings as per \cite{DiffEvol}. The contribution in Algorithm \ref{alg:DE_opt} is the use of the objective  function $f(\tL(x), R) = (1 - E)R$ and its feasible evaluation using MC simulations owing to the high FER property of the ET-QKD channel thus making the joint optimization possible.

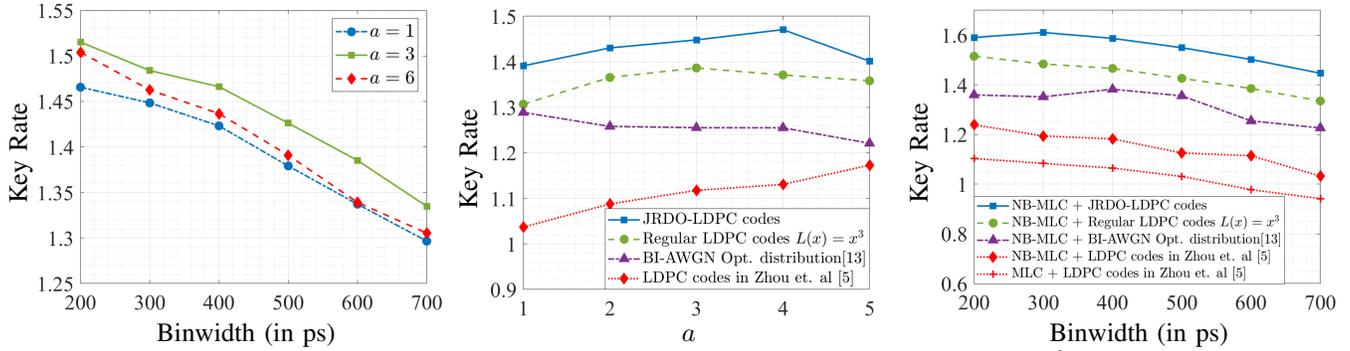
\begin{figure*}[t]
    \centering
    \begin{subfigure}{0.33\linewidth}
\begin{minipage}{0.99\linewidth}
\begin{tikzpicture}
  \node (img) {\includegraphics[scale=0.2]{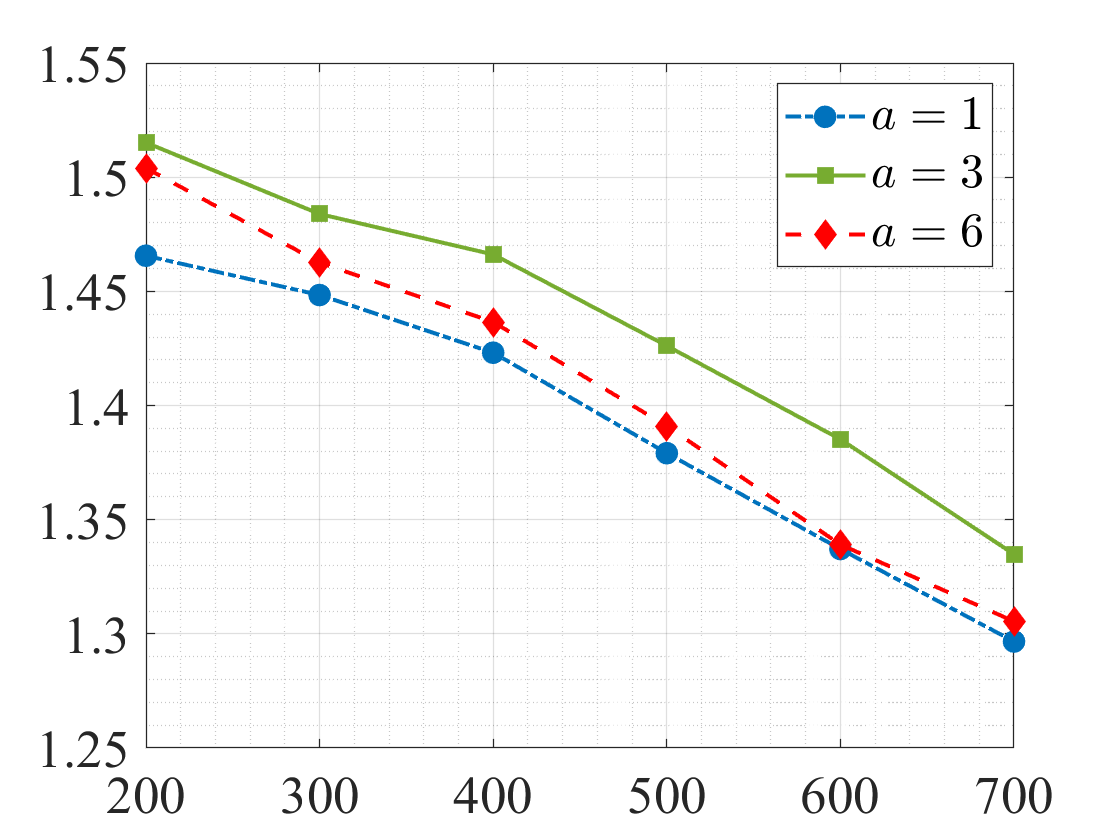}};
  \node[below=of img, node distance=0cm, yshift=1.2cm,font=\color{black}] {Binwidth (in ps)};
  \node[left=of img, node distance=0cm, rotate=90, anchor=center,yshift=-1.1cm,font=\color{black}] {Key Rate};
 \end{tikzpicture}
 \end{minipage}
 \vspace{-3pt}
    \end{subfigure}%
    \begin{subfigure}{0.33\linewidth}
\begin{minipage}{0.99\linewidth}
\begin{tikzpicture}
  \node (img) {\includegraphics[scale=0.2]{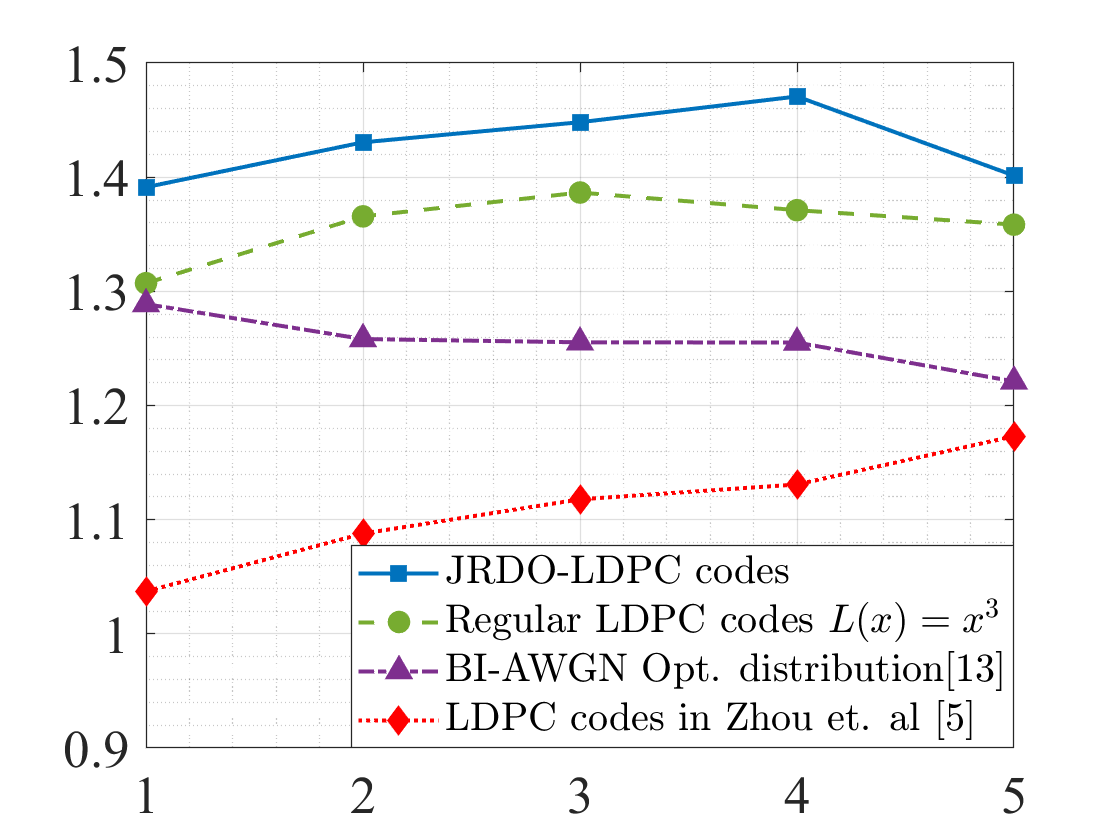}};
  \node[below=of img, node distance=0cm, yshift=1.2cm,font=\color{black}] {$a$};
  \node[left=of img, node distance=0cm, rotate=90, anchor=center,yshift=-1.2cm,font=\color{black}] {Key Rate};
 \end{tikzpicture}
 \end{minipage}
 \vspace{-3pt}
    \end{subfigure}%
        \begin{subfigure}{0.33\linewidth}
\begin{minipage}{0.99\linewidth}
\begin{tikzpicture}
  \node (img) {\includegraphics[scale=0.2]{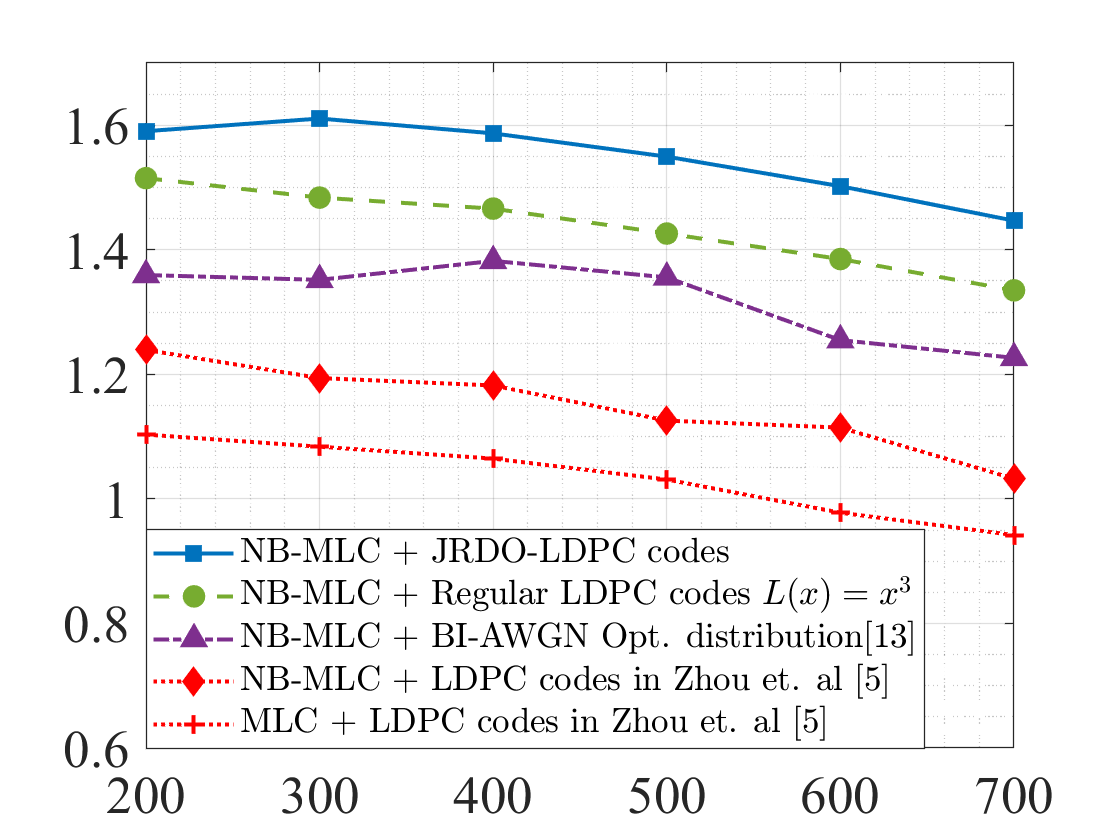}};
  \node[below=of img, node distance=0cm, yshift=1.2cm,font=\color{black}] {Binwidth (in ps)};
  \node[left=of img, node distance=0cm, rotate=90, anchor=center,yshift=-1.2cm,font=\color{black}] {Key Rate};
 \end{tikzpicture}
 \end{minipage}
 \vspace{-3pt}
    \end{subfigure}
     \vspace{-5pt}
    \caption{Left panel: Key rate vs. binwidth for different values of NB-MLC bit size $a$. The QKD system has $2^6$ bins per frame; Middle panel: Key rate vs. $a$ for different LDPC codes. The QKD system has $2^5$ bins per frame and a binwidth of 300ps; Right panel: Key rate vs. binwidth for different LDPC codes. The NB-MLC scheme uses $a=3$. The QKD system has $2^6$ bins per frame.}
    \label{fig:all_codes}
    \vspace{-15pt}
\end{figure*}

\vspace{-0.15cm}
 \begin{algorithm}
\caption{JRDO: Joint Rate and deg. Dist. optimization}\label{alg:DE_opt}
\begin{algorithmic}[1]
\State Initialize population $\Pi = \{(\tL_1,R_1), \ldots, (\tL_{N_{pop},R_{N_{pop}}})\}$

\For{max number of iterations}
\For{$j = 1:N_{pop}$}
 \State $(\tL^{m}_j,R^{m}_j) =$\tDiffMutation$(j, \Pi)$ 
 \State $(\tL^{c}_j,R^{c}_j) = $ \tCrossOver$\big((\tL^{m}_j,R^{m}_j),(\tL_j,R_j)\big)$
 \State Evaluate $f(\tL^{c}_j,R^{c}_j)$ using Monte-Carlo simulations
\EndFor
\For{$j = 1:N_{pop}$}
\If{$f(\tL^{c}_j,R^{c}_j) > f(\tL_j,R_j)$}
   \State Update population: $(\tL_j,R_j) \leftarrow (\tL^{c}_j,R^{c}_j)$
\EndIf
\EndFor

\EndFor

\State \textbf{Output:} $(\tL,R)$ from $\Pi$ with largest $f(\tL,R)$

\end{algorithmic}
\end{algorithm}
\vspace{-0.15cm}

\vspace{-0.2cm}
\section{Simulation Results}\label{sec:sims}
\vspace{-0.1cm}
In this section, we demonstrate the performance of the NB-MLC scheme and the performance of the codes designed using the JRDO algorithm. We compare the performance with the MLC scheme of \cite{MLC} as well as with codes used in \cite{MLC} and codes designed for the BIAWGN channel \cite{DegreeDist}. For the JRDO framework, we optimize degree distribution $\tL(x) = \sum_{d=2}^{5}\tL_d$ where we set $\tL_1$ to be zero and a maximum VN degree of 5. %
For non-JRDO codes, we find and use the rate that results in the largest key rate (for that particular layer of the NB-MLC scheme) by exhaustively iterating over the message length $m_i$. The latency of the NB-MLC scheme is calculated as the sum of the decoding latencies of all the layers in the NB-MLC scheme. We use a code length $N = 2000$ and FFT-based sum-product LDPC decoding (SW version \cite{SW-decoding}) in our simulations. 

In Fig. \ref{fig:a_Varied}, we study the effect of the MLC bit size $a$ on

\noindent
the key rate and latency of the QKD system. Recall that the QKD system has $2^q$ bins per frame implying a total bit size of $q$. %
From Fig. \ref{fig:a_Varied} left panel, we can see that for all values of $q$, the key rate is non-monotonic in $a$ and has a maximum when $a$ is strictly between $1$ and $q$. The reason the key rate is non-monotonic in $a$ is the following.  Increasing the value of $a$ makes the NB-MLC scheme use NB-LDPC codes from a larger Galois field which are stronger resulting in greater FER performance and hence better key rates per layer. However, due to layering, correct decoding in the earlier layers still contributes to the overall key rate even if decoding failures exist in the later layers. This additive effect (due to Eqn. \eqref{eqn:key_rate_mlc}) improves the overall key rate with more layers (small $a$). Due to the above two effects, the overall key rate is non-monotonic. In Fig. \ref{fig:a_Varied} middle panel, we plot the latency of the NB-MLC scheme as a function of $a$ for different values of $q$. From the plot, we see
\deb{that the latency becomes significantly large as $a$ becomes large (close to $q$). This trend is because a larger $a$ implies an NB-LDPC code from a larger Galois field and hence a larger decoding latency.
Note that as $a$ increases, the increase in decoding complexity is sometimes offset by the decrease in the number of decoding layers in the NB-MLC scheme, and thus latency is non-monotonic in $a$ (also evident in Fig. \ref{fig:a_Varied} right panel).}
In Fig. \ref{fig:a_Varied} right panel, we plot the key rate vs. latency achieved due to different values of $a$ in the NB-MLC scheme. From the figure, it is clear that the best trade-off is obtained for a small value of $a$ (3 or 4). Increasing $a$ further results in higher latency at no increase in key rates. 

In Fig. \ref{fig:all_codes} left panel, we compare the key rates across different values of binwidths in a QKD system with $q = 6$. We compare the key rates for $a = 1, 3$, and $6$. Similar to Fig. \ref{fig:a_Varied}, we can again see that across all binwidths, $a = 3$ has a higher key rate compared to $a = 1$ (MLC scheme of \cite{MLC} with binary LDPC codes) and $a = 6$ (a scheme with complete NB-LDPC codes and no layering). Overall, the NB-MLC scheme with a small value of $a > 1$ results in the best system performance. 

In Fig. \ref{fig:all_codes} middle and right panels, we compare the key rates obtained by different code constructions. 
The \deb{diamond marked} curves correspond to LDPC codes used in \cite{MLC}. As per \cite{MLC}, these LDPC codes are randomly constructed such that each VN has a constant degree of 3. Note that there is no limitation on the CN degree distribution in \cite{MLC}. However, the LDPC codes considered in this paper (see Section \ref{sec:ldpc_prelims}) have a two-element CN degree distribution. The \deb{triangle marked} curves correspond to the degree distribution provided in \cite[Table I]{DegreeDist} with a maximum VN degree 5. Note that this degree distribution is optimized for the BIAWGN channel. The \deb{circle marked} curves correspond to LDPC codes with regular VN degree distribution $L(x) = 3$ (similar to \cite{MLC}) but with a two-element CN degree distribution. Finally, the \deb{square marked} curves correspond to degree distributions obtained using the JRDO algorithm. From the figures, we make the following observations. The key rates for the \deb{diamond marked curves} are worse compared to the \deb{circle marked} curves. This trend suggests that it is better to use a two-element CN degree distribution (as done in our paper). The key rates for the \deb{triangle marked} curves (BIAWGN optimized degree distribution) are worse compared to using regular LDPC codes with VN degree 3 (\deb{circle marked} curves). This trend demonstrates that codes optimized for non-QKD channels do not perform well when used for the QKD channel. Finally, in Fig. \ref{fig:all_codes} middle and right panels, we see that JRDO-LDPC codes (\deb{square marked} curves) result in the largest key rates which is because the JRDO-LDPC codes are optimized for the QKD channel. In Fig. \ref{fig:all_codes} right panel, we additionally plot the key rates achieved using the techniques of \cite{MLC} i.e., MLC scheme ($a = 1$) and VN degree 3 regular LDPC codes and no limitation on CN degree distribution (\deb{plus marked} curve). We see that our techniques (\deb{square marked} curve) provide around 40\% improvement in key rates compared to \cite{MLC}.

\vspace{-0.3cm}
\section{Conclusion}\label{sec:conclusion}
\vspace{-0.15cm}
In this paper, we considered the problem of information reconciliation in ET-QKD and proposed a generalization of the multi-level coding (MLC) scheme of \cite{MLC} called NB-MLC that uses NB-LDPC codes. We showed that the NB-MLC scheme offers flexibility in system design in terms of key rate and latency, and the NB-MLC scheme with a small bit size per layer results in the best trade-off between key rate and latency. Finally, we proposed a framework based on different evolution called JRDO that jointly optimizes the rate and degree distribution for the LDPC codes used in the NB-MLC scheme. JRDO-LDPC codes are optimized for the ET-QKD channel and result in a significant improvement in the key rates compared to LDPC codes used in prior work. Ongoing research is focused on optimizing the edge weight distributions (along with JRDO) to further improve the key rates. 

\vspace{-0.2cm}
\section*{Acknowledgement}
\vspace{-0.15cm}
The authors acknowledge \deb{the NSF grant QuIC-TAQS no. 2137984 and NSF grant EFRI-ACQUIRE no. 1741707.}

\end{document}